\documentclass[numberedappendix, appendixfloats]{emulateapj}
\usepackage{natbib}
\usepackage{color}
\usepackage{comment}

\shorttitle{Photoevaporation of Clumps}

\shortauthors{Nakatani {\em et al.}}

\usepackage{bm, mhchem, comment, longtable, amsmath,  amssymb, longtable} 

\newcommand{\smetal}{Z_\odot}
\newcommand{\metal}{Z}
\newcommand{\pc}{\, {\rm pc}}
\newcommand{\Kelvin}{\, {\rm K}}
\newcommand{\unit}[2]{\, {\rm #1} ^{#2}}
\newcommand{\Msun}{\, M_\odot}
\newcommand{\LFUV}{L_{\rm FUV}}
\newcommand{\LEUV}{\Phi_{\rm EUV}}
\newcommand{\HI}{{\rm H\small~I}}
\newcommand{\eq}[1]{\begin{equation}#1\end{equation}}
\newcommand{\eqarray}[1]{\begin{eqnarray}#1\end{eqnarray}}
\newcommand{\Mclump}{M_{\rm c}}
\newcommand{\Rclump}{R_{\rm c}}
\newcommand{\yr}{\, {\rm yr}}
\newcommand{\mdotph}{\dot{M}_{\rm ph}}
\newcommand{\tcr}{t_{\rm cr}}

\newcommand{\nh}{n_{\rm H}}

\newcommand{\e}[1]{\times 10^{#1}}

\newcommand{\fref}[1]{Figure~\ref{#1}}
\newcommand{\tref}[1]{Table~\ref{#1}}
\newcommand{\eqnref}[1]{Eq.(\ref{#1})} 
\newcommand{\secref}[1]{Section~\ref{#1}}

\newcommand{\cm}[1]{ \, {\rm cm}^{#1}}
\newcommand{\kms}{\, {\rm km} \, {\rm s}^{-1}}

\newcommand{\OI}{O{\,\small I}}
\newcommand{\HII}{H{\,\small II}}

\newcommand{\eV}{{\rm \, eV}}
\newcommand{\Mcl}{M_{\rm cl}}
\newcommand{\Rcl}{R_{\rm cl}}
\newcommand{\braket}[1]{\left(#1\right)}

\newcommand{\FEUV}{F_{\rm EUV}}
\newcommand{\FFUV}{G_{\rm FUV}}
\newcommand{\tdiss}{t_{\rm diss}}
\newcommand{\tff}{t_{\rm f}}

\begin{document}

\title{Photoevaporation of Molecular Gas Clumps Illuminated
  by External Massive Stars: Clump Lifetimes and Metallicity Dependence}

\author{Riouhei~Nakatani\altaffilmark{1}
\&
Naoki~Yoshida\altaffilmark{1,2,3}
}

\altaffiltext{1}{Department of Physics, School of Science, The University of Tokyo, 7-3-1 Hongo, Bunkyo, Tokyo 113-0033, Japan}

\altaffiltext{2}{Kavli Institute for the Physics and Mathematics of the Universe (WPI),
       UT Institute for Advanced Study, The University of Tokyo, Kashiwa, Chiba 277-8583, Japan}

\altaffiltext{3}{Research Center for the Early Universe (RESCEU), 
School of Science, The University of Tokyo, 7-3-1 Hongo, Bunkyo, Tokyo 113-0033, Japan}

\email{r.nakatani@utap.phys.s.u-tokyo.ac.jp}

\begin{abstract}
  We perform a suite of 3D radiation hydrodynamics simulations
  of photoevaporation of molecular gas clumps illuminated by external
  massive stars.
  We study the fate of solar-mass clumps and derive their lifetimes
  with varying the gas metallicity over a range of
  $10^{-3} \, \smetal \leq \metal \leq \smetal $.
  Our simulations incorporate radiation transfer of far ultraviolet
  (FUV) and extreme ultraviolet (EUV)
  photons, and follow atomic/molecular line cooling and
  dust-gas collisional cooling. 
  Nonequilibrium chemistry is coupled with the
  radiative transfer and hydrodynamics in a self-consistent manner.
  We show that radiation-driven shocks compress gas clumps to
  have a volume that is set by the pressure-equilibrium with the hot ambient gas.  
  Radiative cooling enables metal-rich clumps to condense
  and to have small surface areas,
  where photoevaporative flows are launched. 
  For our fiducial set-up with an O-type star at a distance of 0.1 parsec, 
  the resulting photoevaporation rate is as small as $\sim 10^{-5} M_{\odot}/{\rm yr}$
  for metal-rich clumps, but is larger for metal-poor clumps that
  have larger surface areas. 
  The clumps are continuously accelerated away from
  the radiation source by the so-called rocket effect,
  and can travel over $\sim$1 parsec within the lifetime. 
  We also study photoevaporation of clumps in a photo-dissociation region.
  Photoelectric heating is inefficient for metal-poor clumps that contain
  a smaller amount of grains, and thus they survive for over $10^5$ years.
  We conclude that the gas metallicity strongly affects the clump lifetime 
  and thus determines the strength of feedback from massive stars in star-forming regions.
\end{abstract}

\section{Introduction}

Giant molecular clouds (GMCs) have masses greater than $\sim 10^4 \Msun$
and are the largest self-gravitating bodies in galaxies.
It has been revealed through molecular line surveys at millimeter and
sub-millimeter wavelengths
that GMCs have highly inhomogeneous substructures and even clumps 
with a variety of sizes $(\sim 0.1-10\pc)$ and masses $(\sim 1-10^3\Msun)$
\citep{1987_Bally,1992_Bertoldi,1993_Blitz,1999_Evans,2000_Williams, 2007_Munoz}.
Typically, most massive clumps 
are sites of star formation in a GMC.
Stars formed in clusters in those clumps
account for a large fraction
of star formation in GMCs.
This suggests that 
evolution of the stars and their interaction
with the associated clumps 
may set the overall star formation rate and efficiency
in GMCs. The evolution and the fate of clumps in a star-forming region 
has been a study of great interest. 

In a GMC, newly formed massive OB stars illuminate the surrounding gas. 
The radiation of the embedded OB stars 
are so intense that the parental GMC
can be destroyed 
in about ten million years \citep[e.g.][]{1980_Blitz, 2005_Stahler}.
Ultraviolet (UV) radiation from the massive stars 
strongly influences the thermal and chemical state of the 
gas. Far-UV (FUV; $6\eV < h\nu < 13.6\eV$) photons
photodissociate molecules in the medium, 
whereas extreme-UV (EUV; $h\nu > 13.6 \eV$) photons
ionize hydrogen atoms.
Around a massive star, there forms an \HII~region surrounded by 
a large photo-dissociation region (PDR).
Typically, the gas in the PDR is heated to $\sim 100 - 1000 \Kelvin$
by grain photoelectric heating,
while the gas in the \HII~region
is heated to $\sim 10^4 \Kelvin$ by photoionization.
The central \HII~region has a sufficiently 
high pressure to drive hydrodynamic shock waves 
in the ambient gas. 
The shocked gas is compressed to become unstable, often to promote
star formation.
Indication of such ``sequential star formation'' \citep{1977_Elmegreen}
has been actually found in local \HII~regions
\cite[e.g.,][]{2005_Deharveng,2008_Deharveng,2010_Miura}
and those in the Large and Small Magellanic Clouds
\citep{2000_Contursi,2000_Rubio,2003_Barba}.

UV radiation emitted by young massive OB stars 
affects the dynamics and thermal/chemical structure 
of nearby clumps in the \HII~region in a highly complicated manner. 
The EUV radiation of massive stars 
heats and ionizes the low-density gas surrounding the clumps.
The high-pressure drives shocks that propagate through
to the cool interior of the clumps. 
The so-called radiation-driven 
implosion \citep[e.g.,][]{1977_Tenorio-Tagle,1980_Klein, 1989_Bertoldi, 1990_Duvert,1994_Lefloch} 
compresses the interior gas
and can trigger star formation inside the clumps 
around newly-born massive OB stars.
After the implosion phase,
a cometary structure develops 
that consists of a dense core 
and a protruding long tail in the direction 
opposite to the radiation source. 
Such cometary-shaped clumps
formed around hot nearby stars
are termed ``cometary globules'' (CGs).
CGs are commonly found in the vicinity of OB
associations \citep[e.g.,][]{1983_Reipurth},
and are thought to be precursors of the Bok globules \citep{1947_Bok}. 
Cometary-structured clumps 
are also found in nearby planetary nebulae such as Rosette nebula,
Helix nebula, and Gum nebula
\citep{1983_Reipurth, 1996_Odell, 2002_ODell, 2007_ODell}.
Infrared observations have shown that 
star formation activities 
might be actually taking place in these cometary globules
\citep{1991_Sugitani,1994_Sugitani} 
and at the bright rims of the globules \citep{1995_Sugitani,1996_Megeath}.

Besides triggering star formation 
and forming cometary structures, 
UV radiation heats the  
gas on the surface of a clump and drive outflows.
This process is referred to as photoevaporation 
\citep[e.g,][]{1989_Bertoldi,1990_Bertoldi}.
Illuminated clumps are gradually eroded from the surface 
and lose a significant fraction of their mass.
Clump photoevaporation 
accompanied by the radiation-driven implosion 
and the subsequent CG formation
has been studied 
both numerically \citep{1977_Tenorio-Tagle, 1980_Klein,1982_Sandford, 1984_Sandford, 1994_Lefloch,1998_Mellema}
and analytically \citep{1989_Bertoldi,1990_Bertoldi, 1998_Mellema},
starting with \cite{1955_Oort}.
Clump photoevaporation models had been directly compared 
with the observed features of CGs in the nearby nebulae.
The theoretical models are largely consistent with the observations
\citep{2001_Lopez-Martin, 2001_Williams}.
Other studies investigate clump photoevaporation due to ionizing radiation 
with self-gravity \citep{2003_Kessel-Deynet,2007_Esquivel}
and with the diffuse EUV \citep{1998_Canto, 2001_Pavlakis}.
Clump photoevaporation of highly inhomogeneous clumps \citep{2005_Gonzalez,2005_Raga}
and of a multi-clump system \citep{2009_Raga}
are also studies with radiation-hydrodynamics simulations.

FUV photons also drive photoevaporation. 
The influences of FUV from external massive stars on
gas clumps (and protoplanetary disks) 
is studied by \cite{1998_Johnstone}
by one-dimensional analytic modeling and hydrodynamics simulations. 
\cite{2002_Gorti} develop a one-dimensional analytic model
of FUV photoevaporation, and explore a wide variety of 
input parameters as a generalization 
of \cite{1998_Johnstone} and \cite{1999_Storzer}.
The authors conclude that strong FUV radiation drives shocks
in an irradiated clump, which 
may trigger star formation inside the compressed core.
The lifetime is
estimated to be $\sim 10^4 - 10^5 \yr$ for $1\Msun$ clumps
with typical parameters in a star-forming region.
More recently, \cite{2017_Decataldo} perform 
1D radiation-hydrodynamics simulations 
with the effects of both FUV and EUV 
to study photoevaporation of clumps 
around massive stars or quasars. 
The derived clump lifetimes 
are in good agreement with those found in \cite{2002_Gorti}.

These previous studies consider clump photoevaporation 
in solar-metallicity environments.
There can be a number of differences in the evolution and the
fate of clumps in low-metallicity environments such as,
for example, star-forming regions in the
LMC \citep{2011_Minamidani} or
in the early universe \citep{2004_Shapiro}.
Radiation cooling by metals and dust  
is inefficient in a low-metallicity clump, 
and the FUV photoelectric heating rate
decreases with the amount of dust grains.
Hence the gas metallicity is likely a critical parameter
that determines 
the thermal/chemical/dynamical structure of clumps. 
Clump lifetimes and the metallicity dependence may be an
important factor that controls
the star formation efficiency in high-redshift galaxies. 
Clearly, it is important to study photoevaporation 
of clumps in various environments.

In the present study, 
we perform 3D radiation hydrodynamics simulations 
of clump photoevaporation by FUV and EUV irradiation.
We solve nonequilibrium chemistry coupled
with all the relevant atomic/molecular line cooling and 
dust-gas collisional cooling. 
Radiative transfer, hydrodynamics, and non-equilibrium chemistry 
are solved in a self-consistent manner, allowing
us to examine clump evolution in detail. 
We vary the gas (clump) 
metallicity over a wider range of
$10^{-3} \, \smetal \leq \metal \leq \smetal$,
and study the fate of the clumps.

This paper is organized as follows.
We describe the methods in \secref{sec:methods}. 
We present the results and discussions
in \secref{sec:results} and \secref{sec:discussion}, respectively.
In \secref{sec:summary}, 
our conclusions and a summary are given.

\section{Methods}
\label{sec:methods}
\begin{figure}[htbp]
\begin{center}
\includegraphics[clip, width=\linewidth]{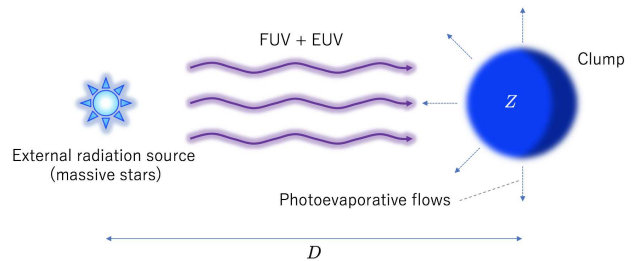}
\caption{A schematic picture of our simulation configuration.
  A clump with metallicity $\metal$ is illuminated by the
  plane-parallel UV radiation 
  from the external source located at distance $D$.
  Photoevaporative flows are excited on the surface of the clump.
}
\label{fig:sche}
\end{center}
\end{figure}

We consider an initially spherical clump with metallicity $\metal$ 
illuminated by an
external radiation source (massive star) 
at distance $D$ (\fref{fig:sche}).
The radiation source is assumed to have 
an FUV luminosity $\LFUV$ and an EUV photon emission rate  
$\LEUV$.
The clump is exposed 
to the FUV and EUV fluxes of 
$\LFUV/4\pi D^2$ and $\LEUV/4\pi D^2$.
We use the publicly available radiation hydrodynamics
code PLUTO \citep{2007_Mignone},
suitably modified for the present study.
In the following,
we describe the numerical methods for our simulations
with giving a brief review of the code.
Further details are found in \cite{2018_Nakatani} (hereafter Paper I) and \cite{2018_Nakatanib},
where the metallicity dependence of protoplanetary disk photoevaporation
is investigated.

We assume that the interstellar medium consists of dust and gas
that contains seven chemical species: 
H, \ce{H+}, \ce{H2}, \ce{C+}, O, CO, e$^-$.
The carbon and oxygen abundances are set to be
$0.926\e{-4} \, \metal/\smetal$ 
and $3.568\e{-4} \, \metal / \smetal$, respectively.
We set the dust-to-gas-mass ratio to be $0.01 \, \metal/\smetal$.
We incorporate heating by \HI~photoionization (hereafter EUV heating)
and photoelectric heating \citep[hereafter FUV heating;][]{1994_BakesTielens}.
FUV and EUV radiative transfer  
are followed at each time step by ray-tracing. 
We do not solve radiative transfer for 
diffuse radiation components.
Diffuse EUV photons
can slightly change the geometrical structure of the cloud
in the early phase of its evolution,
but would not cause significant differences in the cloud evolution,
as demonstrated by \cite{2001_Pavlakis}.
One can also naively expect that diffuse or scattered FUV photons
might change the thermal and/or chemical structure 
especially in shadowed regions, but there, strong photoevaporative
flows are not generated.
Although it would be ideal to include radiation transfer of diffuse
components, we focus, in the present study, on clump photoevaporation
and lifetimes that are largely determined by the effect of direct UV photons
from the radiation source.

In a fully molecular gas cloud that we consider in the present study,
EUV photons can directly ionize \ce{H2} to produce \ce{H2+} at the beginning of the clump evolution.
Since the absorption cross section of \ce{H2} is comparable to that of \HI, 
\ce{H2} photoionization can be more efficient than \ce{H2} photodissociation on the surface of the clump. 
However, the dissociative recombination \ce{H2+ + e -> H + H}
and subsequent \HI~photoionization occur sufficiently rapidly,
so that the produced \ce{H2+} is immediately converted to \ce{H+} to form an \HII~region. 
In order to examine the effects of \ce{H2} photoionization and the associated photoheating, 
we have performed test simulations with these processes. 
We have found that \ce{H2} ionization hardly changes
the evolution of the clumps. 
This allows us to omit the \ce{H2} ionization process in our chemistry model and to save computational cost. 

The simulation volume is defined with 3D cartesian coordinates.
The clump center is initially located at $\vec{r} _ {\rm ini} \equiv (x, y, z) = (0, 0, 0)$.
We set the computational box extending
from $-0.2\pc$ to $1.0\pc$ along the $x$-axis,
and from $0.0\pc$ to $0.2\pc$ along each of the $y$- and $z-$axes,
assuming symmetries with respect to both the $xy$-plane and $xz$-plane. 
The computational grid is uniformly spaced with the number of the cells 
$N_x \times N_y \times N_z = 384\times 64 \times 64$.
The radiation is turned on at $t = 0$, 
and the clump is illuminated by the source
located at $0.1\pc$ from the clump surface.
We adopt a distant-source approximation; 
a plane-parallel UV radiation is incident
on the computational domain at $x = -0.2\pc$.
We set the incident FUV flux and EUV photon number flux
to be $\FFUV =  6.8\e{3} \, G_0$ and
$\FEUV = 5.9\e{12}\unit{s}{-1}{\rm cm}^{-2}$,
where the FUV flux is measured in the unit of the average interstellar flux 
$G_0 = 1.6\e{-3}\unit{erg}{}\unit{cm}{-2} \unit{s}{-1}$.
These fluxes correspond to
the source luminosity of $\LFUV = 1.3\times 10^{37} \unit{erg}{}\unit{s}{-1}$
and $\LEUV = 7.0 \times 10^{48} \unit{s}{-1}$.
In \secref{sec:lumi}, we also consider lower UV fluxes in order to study
the variation of clump lifetimes.
We do not incorporate gravity in our simulations but
refer the readers to our discussion in \secref{sec:gravity}.

We model the initial clump as a Bonner-Ebert (BE) sphere \citep{1955_Ebert,1956_Bonnor}.
The clump is assumed to be initially fully molecular.
We follow \cite{2017_Decataldo}
and adopt $\nh (\vec{r}_{\rm ini})= 2\times 10^4\, \cm{-3}$, $T_{\rm ini} = 10\Kelvin$,
and $P_{\rm c} = 6.9\times 10^{-12} \unit{erg}{} \unit{cm}{-3}$
as the initial hydrogen nuclei density 
and temperature at the clump center,
and confining pressure, respectively.
The corresponding initial clump radius is $R_{\rm ini} = 8.8\e{-2}\pc$,
and the initial clump mass is $M_{\rm ini} = 0.92 \Msun$,
which is smaller than $M_{\rm BE} \simeq 1.18\, c_s^4/\sqrt{P_{\rm c} G^3} \simeq 2 \Msun$.

In order to examine clump photoevaporation 
in an \HII~region and in a photodissociation region (PDR),
and in order to investigate the effects of FUV and EUV separately,
we run three sets of simulations with
(i) both EUV 
and FUV photons,
(ii) only FUV,
and (iii) only EUV.
Hereafter, we label the sets of our simulations 
according to the incorporated processes. 
A set of simulations labeled as ``XX'' specifies which of the photo-heating sources are included. 
For instance, runs with both FUV and EUV are labeled ``FE'',
and runs with only EUV are labeled as ``E''. 
In addition, we indicate the assumed gas metallicity 
by appending ``Z$C$'' to the labels.
This label indicates that  
$\metal = 10^C \, \smetal$ is adopted in the simulation.

\section{Results}
\label{sec:results}
	\subsection{Solar-Metallicity Clumps}
	\label{sec:solar}

        \subsubsection{Photoevaporation and clump lifetime}
        We first focus on the simulation result for a solar-metallicity clump. 
	When the radiation source is turned on,
	the incident FUV photons 
	quickly dissociate \ce{H2} molecules 
	in the entire hemisphere  
	facing the source. 
	The incident EUV photons then immediately 
	ionize the hydrogen atoms 
	produced by the photodissociation. 
	The photo-chemical reactions yield three distinct
	regions: a molecular core region, 
	a geometrically thin \HI~shell with
        a typical thickness of $\sim 0.01 \pc$,
	and an \HII~region surrounding the neutral regions.
	In the neutral regions, 
	a bowl-shaped shock is driven, which propagates
        through the interior of the clump and compresses the gas there.
	This is the so-called radiation-driven implosion phase \citep{1994_Lefloch}.
	Hereafter, we simply refer to the compression due to the shock
        as shock-compression. 
		
\begin{figure*}[htbp]
	\begin{center}
	\includegraphics[clip, width = \linewidth]{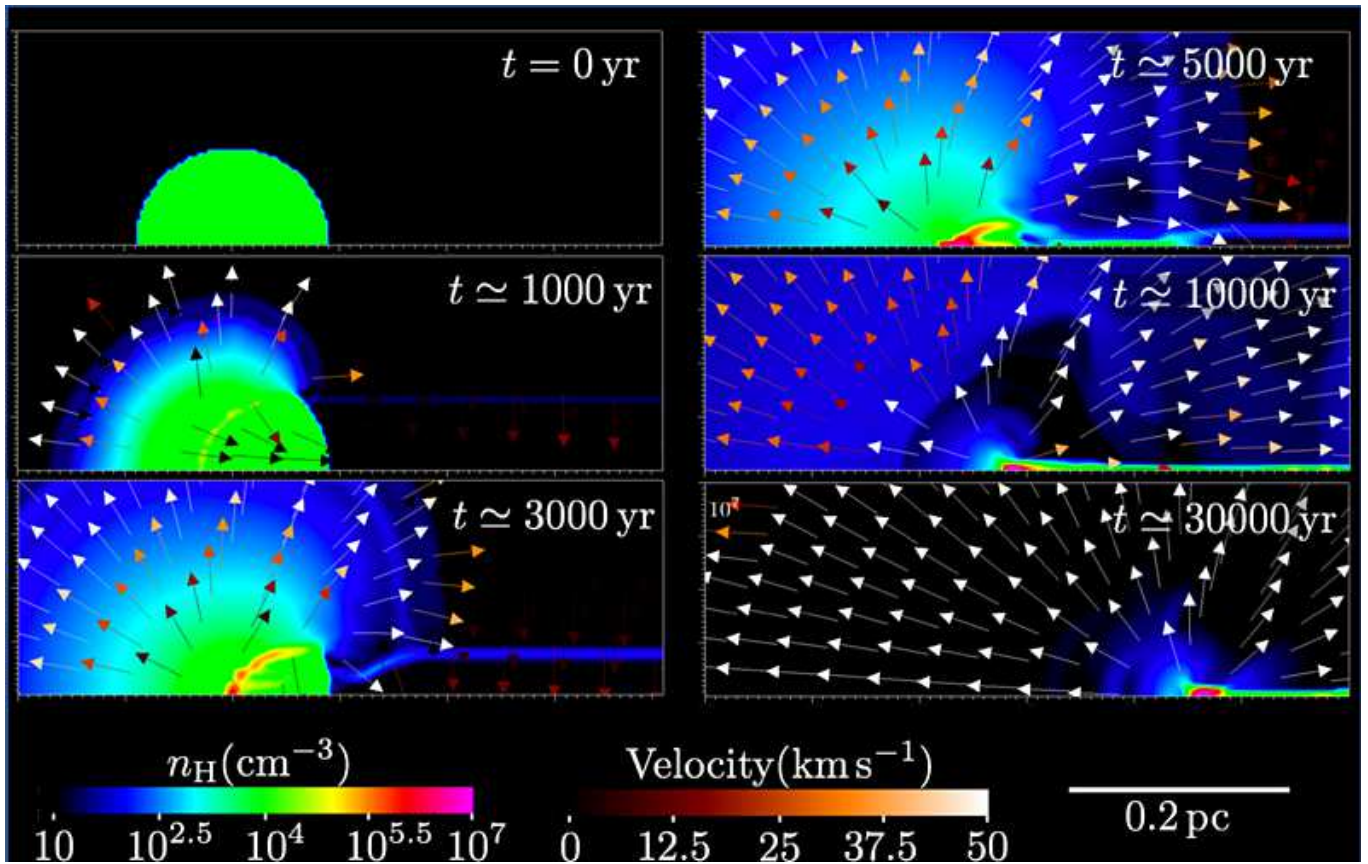}
	\caption{Time evolution of the 
	solar-metallicity clump
	illuminated by FUV/EUV photons.
	The color maps show the cross-sectional density distribution.
	The velocity field is indicated by the arrows 
	colored according to the magnitude.
	The physical length scale is indicated at the bottom right. 
	The radiation source is placed at the left of the computational domain.}
	\label{fig:simz0}
	\end{center}
	\end{figure*}

	The shock propagates through the clump
	and converges toward the $x$-axis
	(see the snapshots at $t \simeq 1000\text{--}5000\yr$ in \fref{fig:simz0}).
	The gas cools quickly after the shock passes through, 
	because the characteristic timescales
        of \OI~cooling and dust-gas collisional cooling
	are much shorter than the shock crossing time.
	Thus the clump temperature remains being low at $T \simeq 10\Kelvin$.
	The low temperature and the low ionization degree $(< 10^{-5})$ in the neutral region
	assure that \OI~remains
	effective as a major coolant during the implosion phase
	without being destroyed through the rapidly-proceeding charge transfer 
	reaction \ce{H+ + O -> H+ + O} \citep{2011_Draine}.	
	Owing to the low temperature (hence low pressure),  
	the clump is compressed by the external hot ionization front,	
	even after the shock converges towards the $x$-axis.
    	The clump is continuously compressed 
	until the internal density becomes 
	high enough to yield comparable internal pressure 
	with the external pressure.
	This approximate pressure balance is achieved 
	from the portion closer to the radiation source.
	The shock-compression and the external pressure 
        raise the average density of the neutral clump in the first $\sim 7000 \yr$.
	After this compression phase, 
	the clump expands slightly 
	and generates
	a cometary structure in the rest frame of the clump
	with the typical internal density of $\nh \sim 10^7\cm{-3}$
	(\fref{fig:simz0}).
	
	The UV radiation drives 
	photoevaporative winds (\fref{fig:simz0}),
        which are launched from the ionization front
	at a velocity of $\simeq 10 - 30 \kms$ with $\nh \sim 10^5 \cm{-3}$, 
	stripping the mass at the wind base. 
	Throughout the present study,
	we calculate the clump mass $\Mclump$ as 
	\eq{
		\Mclump	=	\int _{V(\Rclump, \bm{r}' )} \rho \, dV,	\label{eq:clumpmass}
	}
	where $V(\Rclump, \bm{r}')$
	is the volume of a sphere with radius $\Rclump$ from the center position $\bm{r}'$.
	We use $R_{\rm ini}$ for $\Rclump$,
	while we set $\bm{r}'$ to be the position of 
	the center of mass in the ``dense'' region defined as
	\eq{
		\nh >  n_{\rm d} \equiv {\rm min} \left(\frac{\nh (\vec{r}_{\rm ini})}{2}, \,\frac{n_{\rm max}}{2}\, \right),
	}
	where $n_{\rm max}$ is the maximum 
	hydrogen nuclei density in the computational domain at each time step.
	We adopt this definition
	to estimate the mass for the clumps in all the FE, F, and E runs.
	
\begin{figure}[htbp]
	\begin{center}
	\includegraphics[clip, width = \linewidth]{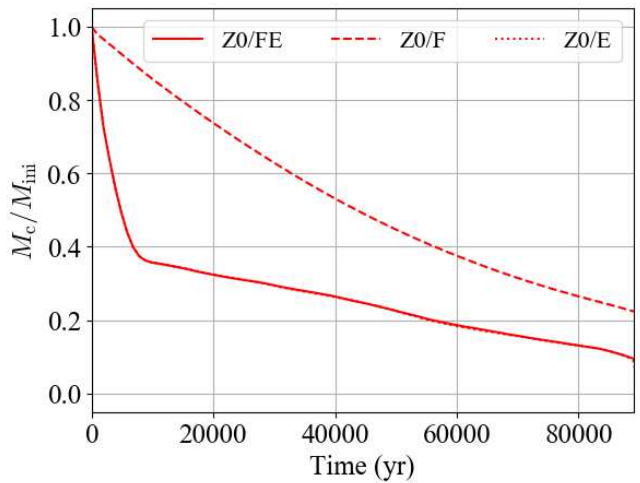}
	\caption{The clump mass ratio $\Mclump/M_{\rm ini}$ as a function of time.
	The solid, dashed, and dotted lines show the results for 
	Z0/FE, Z0/F, and Z0/E, respectively. 
	The solid and dotted lines almost overlap.} 
	\label{fig:Mt}
	\end{center}
	\end{figure}

	\fref{fig:Mt} shows the ratio of the clump mass $\Mclump$ to the initial clump mass 
	$M_{\rm ini} (= 0.92 \Msun)$ as a function of time.
	The clump mass monotonically decreases
	in all the cases of Z0/FE, Z0/F, and Z0/E.
	In run Z0/FE 
	the clump loses about a half of its initial mass
	in the implosion phase	%
	in the first $\sim 10^4$ years.
	The lost mass was contained in the part where
	the R-type ionization front sweeps 
	before it converts to D-type
	\citep{1978_Spitzer,2004_Shapiro}.
	The ionized gas evaporates and flows outward from the clump center, 
	over approximately a crossing time of the EUV-driven flows
        of $R_{\rm ini}/10\kms \sim 10^4 \yr$. 
	In run Z0/F, 
	FUV photons heat nearly the entire hemisphere facing to the radiation source
	to temperatures of $\sim 300 - 500 \Kelvin$.
	The heated region starts to expand
	at a velocity of $1-3 \unit{km}{}\unit{s}{-1}$.
	The outflow is significantly slower than the EUV-driven flows.
	Therefore, 
	the dispersal time of the clump 
	is longer than 
	in Z0/FE or Z0/E (\fref{fig:Mt}).
	\cite{2002_Gorti} show that 
	the evolution of a clump which is impulsively illuminated by 
	FUV radiation bifurcates, 
	depending on the effective optical depth parameter $\eta_0$ defined as
	\eq{
		\eta_0 = \frac{R_{\rm ini}n_0}{N_0},	\label{eq:etazero}
	}
	where $n_0$ is the initial internal density of the clump
	and $N_0 \sim 2\e{21} \cm{-2}$ is the column heated by FUV.  
	If $\eta_0 > 1$, the FUV-heated region is confined to 
	a thin surface layer of the clump.
	A shock is driven inside the clump 
	when FUV heating is sufficiently strong. 
	If $\eta_0 < 1$, 
	the clump is entirely heated and expands 
	on the crossing timescale.
	Our clump with $\eta_0 \simeq 1$ 
	has an intermediate evolutional character:
	FUV heats almost entire region of the clump.
	Especially, the hemisphere facing to the radiation source is heated to high temperatures. 
	FUV-driven flows are quickly excited,
	and weak shocks are generated inside the clump. 
	The clump eventually forms a cometary structure at $t \sim 7\e{4} \yr$.

	\begin{figure}[htbp]
	\begin{center}
	\includegraphics[clip, width = \linewidth]{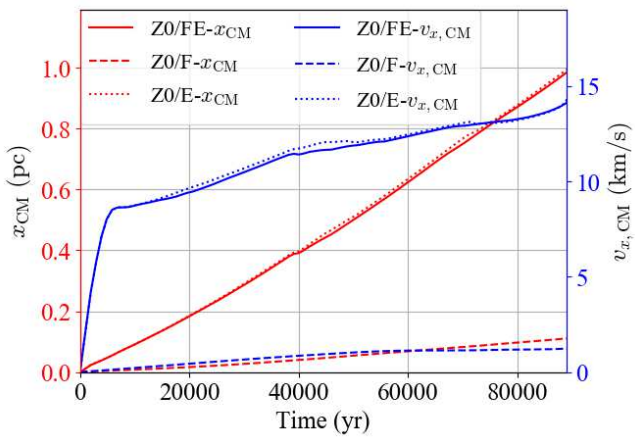}
	\caption{
	The position of the mass center of the ``dense'' region 
	along the $x$-axis $(x_{\rm CM} = x'$; red) and 
	the bulk velocity of the region ($v_{x, \rm CM}$; blue)
	as functions of time.
	The solid, dashed, dotted lines represent 
	those of Z0/FE, Z0/F, and Z0/E, respectively. 
	Note that the solid and dotted lines largely overlap. 
	}
	\label{fig:tvsx}
	\end{center}
	\end{figure}

        \subsubsection{The rocket effect}
	Photoevaporating clumps move away
	from the radiation source 
	by the rocket effect \citep{1955_Oort}.
        It is observed in our simulations that
        clumps are accelerated and move along the $x$-axis 	
	(\fref{fig:simz0} and \fref{fig:tvsx}).	
	The rocket effect can be described in terms of 
	momentum conservation 
	between the clump and the materials ejected
	by photoevaporation.
	Assuming that 
	photoevaporative flows are launched 
	with density $\rho_{\rm b}$ and velocity $v_{\rm b}$	
	from the surface of the hemisphere facing the radiation source,
	we can calculate the acceleration as 
	\eqarray{
	 \frac{dv_{\rm cl}}{dt} &=& \frac{1}{M_{\rm cl}}
	\left( \frac{1}{2} S \rho_{\rm b} v_{\rm b}^2\right) \label{eq:rocket2} \\
	&\simeq &  4.0\e{-13} \unit{km}{}\unit{s}{-2} 
	\left(\frac{S}{10^{-4} \pc ^2}\right)
	\left(\frac{n_{\rm b}}{10^5 \cm{-3}}\right) 			\nonumber \\
	&&
	\times \left(\frac{v_{\rm b}}{10 \unit{km/s}{}}\right)^2
	\left(\frac{M_{\rm cl}}{M_\odot } \right)^{-1} ,	
	\label{eq:rocket}
	}
	where 
	$S$ and $M_{\rm cl}$
	are the launch area of the photoevaporative flows
	and the compressed clump mass. 
	The base hydrogen nuclei density $n_{\rm b}$
	is calculated as $n_{\rm b} \simeq \rho_{\rm b}/m_{\rm H}$,
	where $m_{\rm H}$ is the mass of a hydrogen atom.
	Note the factor of a half in \eqnref{eq:rocket2} that 
	accounts for the net momentum in the $x$-direction
	gained through photoevaporation. 
	We use the values 
	of the clump in our Z0/FE run: 
	$S \simeq 0.8 \e{-4}\pc ^2$, 
	$n_{\rm b} \simeq 1.0\e{5} \cm{-3}$,
	$v_{\rm b} \simeq 20 \kms$,
	and $M_{\rm cl} \simeq 0.38\,\Msun$.
	Substituting these
	into \eqnref{eq:rocket},
	we estimate the acceleration to be
	$a_{\rm cl} = 3\e{-12} \unit{km}{}\unit{s}{-2}$.
	In run Z0/FE,
	the position of 
	the center of mass in the ``dense'' region
	is well approximated by a function of time as
	$x' = a t^2 + b t$ 
	with $a = 1.1 \e{-12} \unit{km}{}\unit{s}{-2} (=3.6\e{-5} \unit{km}{}\unit{s}{-1} \unit{yr}{-1})$
	and $b = 7.6 \unit{km}{}\unit{s}{-1}$	
	(the solid and dotted lines in \fref{fig:tvsx}),
	after the implosion phase $(t \gtrsim 10^4\yr)$.
	The acceleration $a$ is consistent with that estimated by \eqnref{eq:rocket}.
	The coefficient $b$ corresponds to 
	the clump velocity at 
	the end of the implosion phase. 
	In summary, 
	the solar metallicity clump starts to move away from the external radiation source 
	at a velocity of $\sim 10 \unit{km}{}\unit{s}{-1}$
	after the implosion phase, 
	and then continuously accelerated by the rocket effect caused by EUV photoevaporation
	at a rate of $\sim 10^{-5} \unit{km}{}\unit{s}{-1} \unit{yr}{-1}$;
	the clump travels over a distance of $\sim 1\pc$ in $\sim 10^5\yr$,
        when the mass decreases to 10 percent of the initial mass.

\subsection{Photoevaporation of Low-Metallicity Clumps}

	\subsubsection{Effects of EUV radiation}
	\begin{figure*}[htbp]
	\begin{center}
	\includegraphics[clip, width = \linewidth]{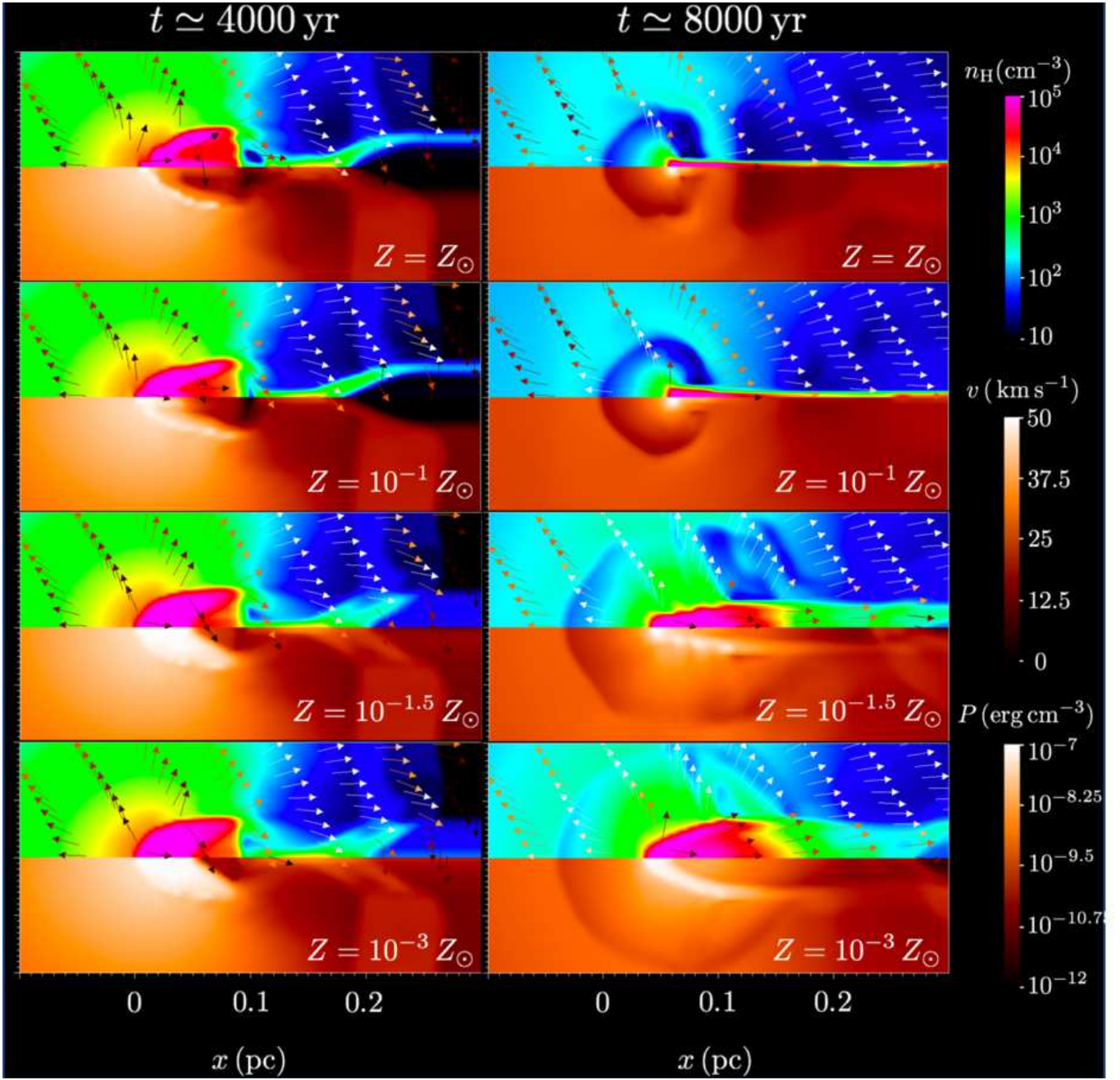}
	\caption{
	We show the density, velocity, and pressure distributions 
	on the $xy$-plane.
	The upper half and the lower half in each panel
	show the distribution of density and pressure, respectively.
	The velocity field is indicated by the arrows colored by the magnitude. 
	The left and right column show the distributions at $t \simeq 4000\yr,~8000\yr$,
	respectively. 
	Note that the density scale is different from that in \fref{fig:simz0}
	for clarity.
	}
	\label{fig:configs}
	\end{center}
	\end{figure*}
\begin{figure}[htbp]
	\begin{center}
	\includegraphics[clip, width = \linewidth]{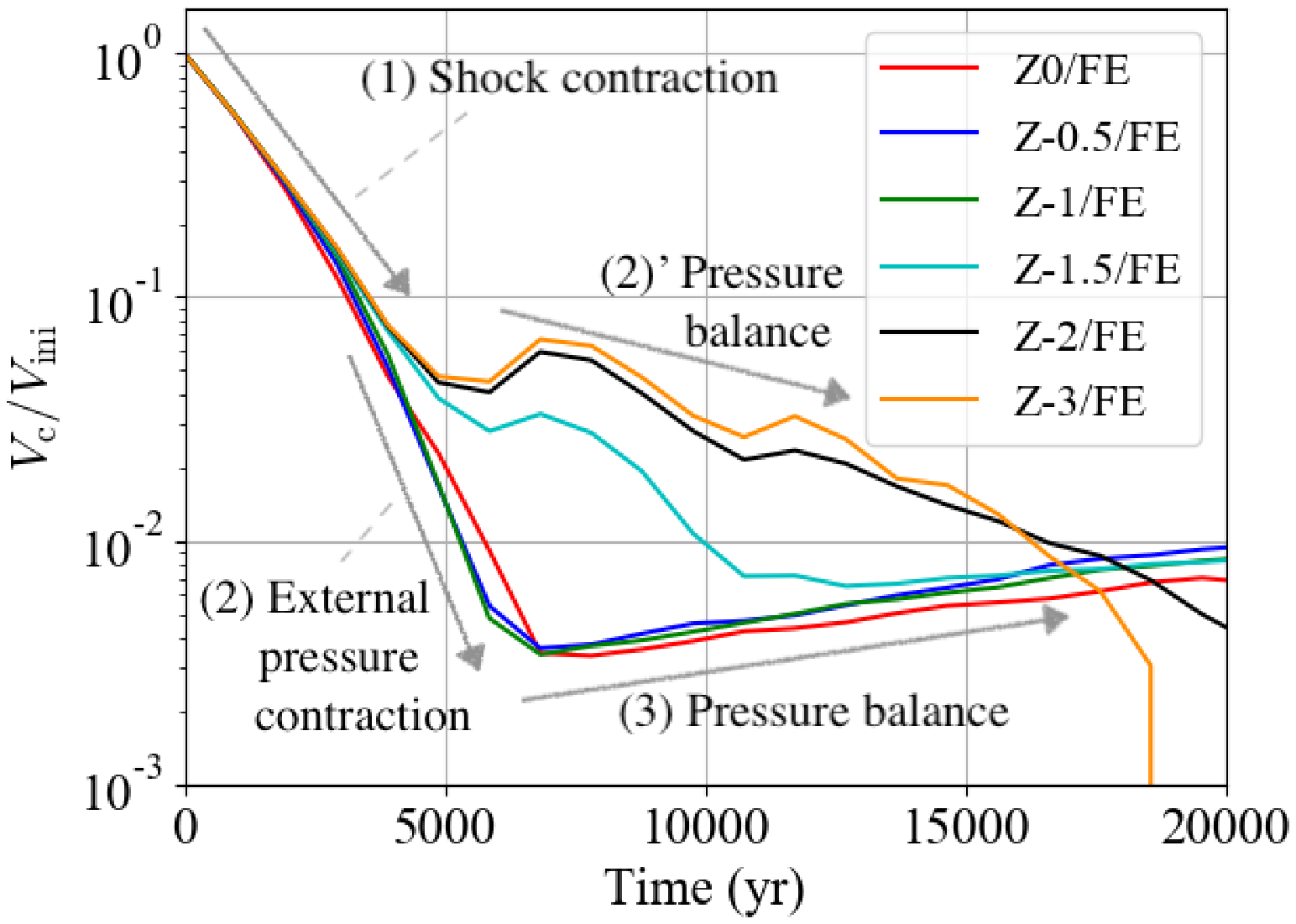}
	\caption{
	The time evolution of the clump volume $V_{\rm c}$ (\eqnref{eq:volume}) for clumps 
	in our FE runs with different metallicities.
	We normalize $V_{\rm c}$ by the initial volume $V_{\rm ini}$.
	}
	\label{fig:mvevolution}
	\end{center}
	\end{figure}

	The initial evolution of low-metallicity clumps is similar to that
        of the solar-metallicity clump.
	The ionization front first develops within
	the radius of $\sim  0.4 \, R_{\rm ini} \simeq 0.04\pc (\equiv R_{i, {\rm ini}})$ 
	from the clump center in the hemisphere facing to the source
	(cf. \fref{fig:simz0}).	
	A bowl-shaped shock (the left column of \fref{fig:configs}) 
	is generated in the neutral region inside the ionization front.
	Clumps 
	are first compressed by the shock
	for $\sim 4000\yr$. 
	The shock-compression time 
	corresponds to the timescale 
	with which the shock propagates through the clump 
	$\tcr = R_{i, {\rm ini}}/ 10 \unit{km}{}\unit{s}{-1} \simeq 4 \e{3} \yr$
	(the phase (1) in \fref{fig:mvevolution}).
	The strong shock with a very large mach number of $\sim 30$
	can heat the post-shock gas
	to a temperature of $\sim 10^4 \Kelvin$.
	This is sufficiently high for the internal pressure
        of the clump to
        equal the external pressure.
	For $\metal \leq 10^{-2} \,\smetal$ clumps,
	radiative cooling processes are not effective owing to the smaller
        amount of metals and dust.
	The internal temperatures reach $\sim  10^3 - 10^4 \Kelvin$ through
        the shock compression,
	yielding comparable pressure between the inside and outside of the clump,
        and the clumps do not shrink.
	Hence, the lower metallicity clumps keep larger volumes 
	than in run Z0 even after the implosion phase
	(the phase (2)' and (3) in \fref{fig:mvevolution}).
        
	The cooling time 
	of a clump with $\metal \gtrsim 10^{-1}\, \smetal$
	is much shorter than the shock propagation time.
	Thus the neutral region cools quickly to below $100 \Kelvin$,
        allowing the clump 
	to be compressed further until 
	the internal density becomes sufficiently high to 
	match the external pressure.

	The thermal energy is lost by 
	line emission and dust-gas collisional heat transfer.
	In particular,
	\OI~cooling and dust-gas collisional cooling 
	are dominant in the neutral region.
	Since the amounts of oxygen and dust are proportional to 
	metallicity,
	the specific \OI~cooling rate
	and dust-gas collisional cooling rate
	increase with metallicity.
	\fref{fig:mvevolution} shows
	the volume of the ``dense'' region
	\eq{
		V_{\rm c}  = \int_{V(\Rclump, \bm{r}')} \, {\rm d}V,	\label{eq:volume}
	}
	as a function of time for each metallicity clump. 
	In the early stage of the clump evolution 
	($t \lesssim \tcr$; the phase (1) in \fref{fig:mvevolution}),
	the characteristic time for \OI~cooling and dust-gas collisional cooling 
	is evaluated as
	\eq{ 
		\begin{split}
		t_{\rm c,OI} &\sim 10^2 \text{--} 10^3   \braket{\frac{\metal} { \,\smetal}}^{-1} \yr \\
	 	t_{\rm c,dust} &\sim 10^2 \text{--} 10^3  \braket{ \frac{\metal} {\,\smetal}}^{-1} 
					\braket{\frac{\nh}{10^4 \, \cm{-3}}}^{-1} \yr,
		\end{split}
		\label{eq:coolingtimes}
	}
	respectively.
	Clearly, the cooling time is sufficiently short for 
	the clumps with $\metal \gtrsim 10^{-1}\,\smetal $,
	compared to $\tcr$.
	Thus,
	the temperature of the inner region of the clump 
	with $\metal \gtrsim 10^{-1} \, \smetal$ is 
	coupled with the dust temperature of $T_{\rm d} \sim 10 \Kelvin$
	at $t \lesssim 1.5\e{4}\yr$.
        	
	\begin{figure*}[htbp]
	\begin{center}
	\includegraphics[clip, width = \linewidth]{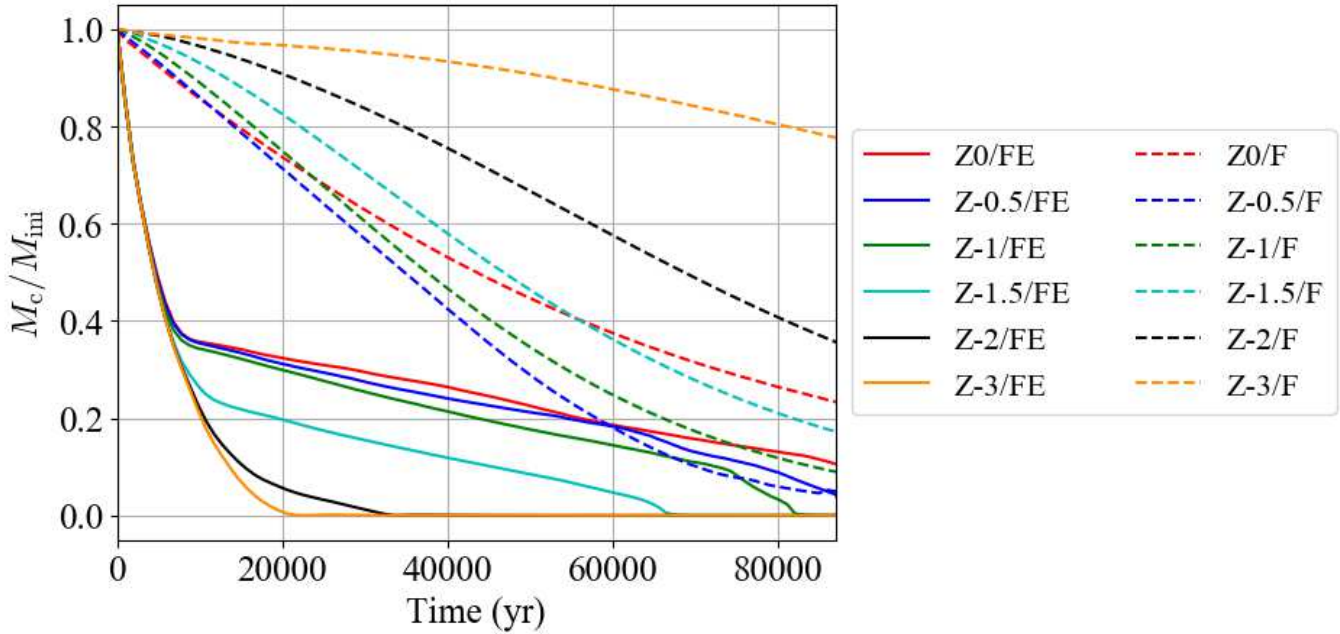}
	\caption{Time evolution of the clump mass with various metallicities. 
	The solid and dashed lines 
	represent the mass evolution in our runs with FUV/EUV and with FUV only,
        respectively. 
	The mass is normalized by the initial clump mass. }
	\label{fig:zvsm}
	\end{center}
	\end{figure*}
	About a half of the initial mass is lost during the implosion phase
	in the FE runs (solid lines \fref{fig:zvsm}; 
	see also the discussions in \secref{sec:solar}).
	In the later phase,
	the gas is gradually stripped off
	from the clump surface through EUV photoevaporative flows.
	The mass loss rate due to EUV photoevaporation is 
	approximated by 
	\eq{
	\mdotph  \simeq \rho_{\rm b} v_{\rm b} S.	\label{eq:prate}
	}
	Since EUV photons are mainly absorbed by hydrogen atoms, 
	$\rho_b$ is independent of metallicity.
	The launch velocity of the EUV-driven photoevaporative winds is 
	of the order of $c_s$, which is
	typically $\sim 10\unit{km}{}\unit{s}{-1}$,
	and is also independent of metallicity.
	Therefore, the metallicity dependence of the mass loss rates
	is caused by the difference in the area $S$ of the evaporating region.

	For the clumps with $\metal \gtrsim 10^{-1} \, \smetal$,
	the compression is the main channel to decrease $S$
	until the compression phase ends at $t \sim 6000\text{--}7000\yr$,
	forming the cometary structure (\fref{fig:configs}).
	After that, 
	the channel is replaced by photoevaporation.
	The mass loss rate $\dot{M}$ is switched to be a smaller value 
	(the red, blue, and green solid lines in \fref{fig:zvsm}).
	For $\metal \lesssim 10^{-2} \, \smetal$, 
	the clump volume hardly decreases in the implosion phase
	because cooling is inefficient to reduce the temperature in the post-shock region.
	The surface area $S$ decreases only slightly owing to photoevaporation
	even in the implosion phase, in contrast to the clumps 
	with $\metal \gtrsim 10^{-1}\, \smetal$.
	Thus, the resulting photoevaporation rate remains large,
	and the clump evaporates in a shorter time than the cases with higher metallicities.
	For $\metal \sim 10^{-1.5} \,\smetal$,
	the cooling time is comparable to $\tcr$ in the beginning,
	but becomes shorter as the density increases by shock-compression
	(cf. \eqnref{eq:coolingtimes}).
	The volume of the clump remains relatively large for a long time
	compared to those with $\metal \gtrsim 10^{-1}\,\smetal$ (\fref{fig:mvevolution});
	the amount of mass lost before forming cometary structure 
	is larger for the $\metal = 10^{-1.5}\, \metal$ clump
	than the $\metal \gtrsim 10^{-1}\,\smetal$ clumps (\fref{fig:zvsm}). 
	Afterwards, 
	the mass loss rate is largely the same as that of the $\metal = 10^{-1}\,\smetal$ clump. 
	The clump evolves in an intermediate manner between  
	the higher metallicity ($\metal \gtrsim 10^{-1}\, \smetal$) clumps 
	and the lower metallicity ($\metal \lesssim 10^{-2}\,\smetal$) clumps.

	We define the lifetime of a clump as 
	the time at which the clump mass 
	decreases to 10 percent of its initial mass.
	\begin{figure}[htbp]
	\begin{center}
	\includegraphics[clip, width = \linewidth]{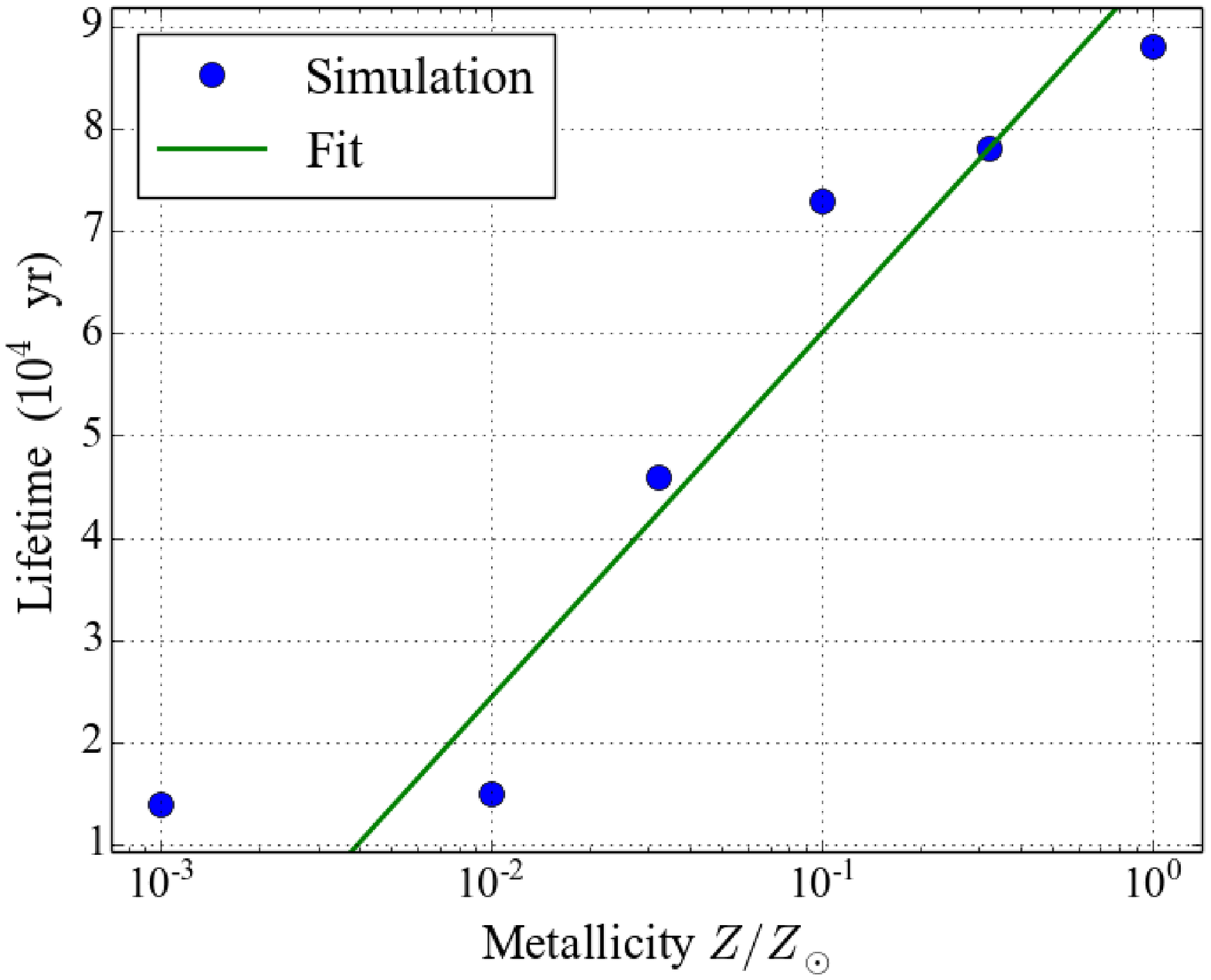}
	\caption{	 The blue dots show the clump lifetimes
			derived from our FE runs (\fref{fig:zvsm}).
			The green line is a fit 
			given as $T_{\rm life}  = [9.6 + 3.6 \log (\metal / \smetal)] \e{4} \yr$
			for $10^{-2} \, \smetal \leq \metal \leq \smetal$.
			}
	\label{fig:lifetimes}
	\end{center}
	\end{figure}
        We find that
	the lifetime decreases with metallicity 
	at $10^{-2} \, \smetal \leq \metal \leq \smetal$ in the FE runs.
	The metallicity dependence is fitted as

        \begin{equation}
	  T_{\rm life}  = [9.6 + 3.6 \log (\metal / \smetal)] \e{4} \yr.
        \end{equation}
        
	The lifetimes is
	roughly constant at $T_{\rm life} \simeq 1.4 \e{4} \yr$ for $\metal \leq 10^{-2} \, \smetal $.
	Metal-rich clumps illuminated 
	by the EUV radiation have longer lifetimes because of the smaller clump sizes
	resulting from the efficient cooling and contraction.

	In order to separate the effects of FUV and EUV in clump photoevaporation,
	we have also performed simulations without FUV radiation (Runs E).
	We have found that the overall evolution is quite similar to
	the FE runs.
	We conclude that 
	the mass loss of clumps is mainly driven by EUV photoevaporation
	at any metallicity,
	and that FUV radiation does not 
	cause a significant impact on the dynamical evolution
	of the clumps. These conclusions hold when a clump is illuminated
        by direct EUV photons, i.e., when the clump is located within an \HII~region.

	\subsubsection{FUV-driven photoevaporation}	
	FUV photons can penetrate through to the outer region of an \HII~region.
	There, in the so-called photodissociation regions (PDRs), 
        molecules are dissociated and the gas is heated by 
	photoelectric heating.
	Since PDRs generally have much larger volumes than 
	the inner/central \HII~regions, 
	there may exist numerous molecular clumps in PDRs.
	It is thus worth studying clump photoevaporation in such regions.
	To this end, we perform an additional set of simulations  
	without EUV radiation. We dub these runs with ``F''. 

        With our fiducial configuration (\secref{sec:methods}),
	a clump with $\metal = 1 \, \smetal$
	is marginally optically thick to the FUV radiation. 
	FUV photons are attenuated in the hemisphere 
	facing the radiation source, 
	where the gas temperature is $300- 500\Kelvin$.
	Photoevaporative flows are driven
	with a typical launch velocity of $1-3 \unit{km}{}\unit{s}{-1}$. 
	Weak shocks are excited inside the clump, 
	and the clump eventually forms a cometary shape
	in a similar manner to \fref{fig:simz0}.

        For $\metal \leq 10^{-0.5} \, \smetal$,
	the clumps are effectively optically thin to FUV,
	and the whole volume of a clump is photo-heated. 
	The FUV photoheating raises the gas temperature to 
	$T_{\rm clump, F} \simeq 200, 150, 100, 30, 10\Kelvin $
	for the clumps with
	$\metal  = 10^{-0.5}, 10^{-1}, 10^{-1.5}, 10^{-2}, 10^{-3}\, \smetal$, respectively.
	The clumps have $\eta_0' < 1$,
	where we generalize
	$\eta_0$ of \eqnref{eq:etazero} to incorporate metallicity dependence as 
	\eq{
		\eta_0 ' (\metal) = \eta_0 \,\metal/\smetal.
	}	
	The dynamical evolution of the clumps in a PDR is largely consistent
	with the study of \cite{2002_Gorti}.
	The clumps expand 
	at the velocity of the order of the sound speed,
	keeping the initial spherical shape. 
	Hydrodynamics shocks are not generated in the low-metallicity clumps.

	We find that the mass loss rate is larger for higher metallicity clumps
	in the range $\metal \leq 10^{-0.5} \, \smetal$,
	owing to the higher $T_{\rm clump, F}$ (the dashed line in \fref{fig:zvsm}).
	The solar-metallicity clump has a smaller evaporation rate than 
	the clumps with sub-solar metallicities,
	because photoevaporation occurs only in the hemisphere 
	facing the external radiation source.
	We note also that the small mass loss rate in a very low-metallicity case with
        $\metal = 10^{-3} \, \smetal $ 
        is likely caused because we do not include self-gravity in our simulations.
	We discuss the effect of self-gravity in \secref{sec:gravity}.


\section{Discussions}	 \label{sec:discussion}

\subsection{Clump Photoevaporation with Lower UV fluxes}
\label{sec:lumi}
        We examine possible variations of our results
        due to different assumptions and to details of the simulation set-up. 
	We first study the effect of UV flux. To this end, we perform simulations with 
	smaller incident UV fluxes of 
	$(\FFUV, \FEUV) = ( 8.4\e{2}\,G_0, \, 2.1\e{11}\unit{s}{-1}),~
	(5.7\e{1}\,G_0, \, 2.5\e{9}\unit{s}{-1})$.
	This also allows us to make comparison with the study of, e.g., \cite{2017_Decataldo}.
        We refer to these parameter sets 
	as the intermediate and weak cases, 
	respectively,
	and differentiate them from the fiducial one.
	\begin{table*}[htp]
	\caption{UV luminosities used in the simulations.}
	\begin{center}
	\begin{tabular}{l | c | c c | c c}  
	Case	&	
	Label& 
	 $\FFUV\,(G_0)$& 
	 $\FEUV\,(\cm{-2}\unit{s}{-1})$	&
	$\LFUV$ $(\unit{erg}{}\unit{s}{-1})$&	
	 $\LEUV$ $(\unit{s}{-1})$		 \\ \hline 
	Fiducial 	&	H&$ 6.8\e{3}$	& $5.9\e{12}$	&	$1.3\e{37} $	& 	$ 7.0 \e{48} $	\\
	Intermediate	&M&$ 8.4\e{2}$	& $2.1\e{11}$&	$1.6\e{36}$	&	$2.5 \e{47} $	\\
	Weak	&	L& $5.7\e{1}$	& $2.5\e{9}$&	$1.1\e{35}$	&	$3.0 \e{45}$	\\	
	\end{tabular}
	\end{center}
	\label{tab:uv}
	\end{table*}%
	In \tref{tab:uv},
	we list the sets of the fluxes 
	and the corresponding 
	luminosities of the external radiation source 
	which is assumed to be located at $0.1 \pc$ from the clump surface.
	Hereafter in this section, we specify a simulation by naming it as, e.g.,
	``Z-0.5/F(M)''.
	The letter in the parenthesis indicates the labels in \tref{tab:uv}.

\begin{figure}[htbp]
	\begin{center}\includegraphics[clip, width = \linewidth]{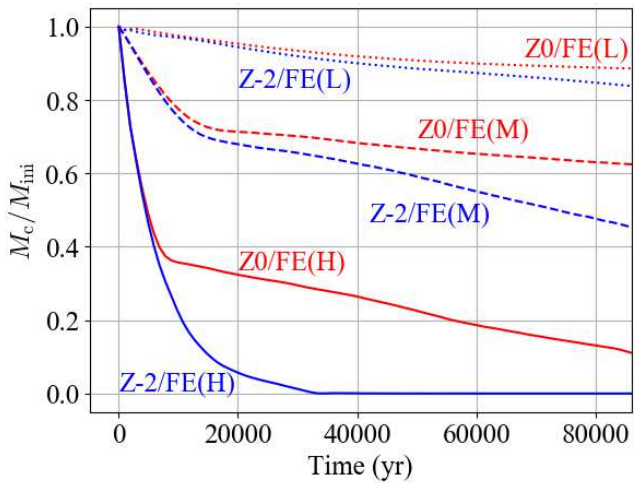}
	\caption{The time evolution of the clump mass relative to the initial mass
	$\Mclump/M_{\rm ini}$ for Z0/FE and Z-2/FE
	with each of the luminosity sets.
	The solid, dashed, and dotted line indicate
	the simulations with the fiducial (H), 
	intermediate (M), and weak (L) luminosity sets, respectively. 
	}
	\label{fig:LdepFE}
	\end{center}
	\end{figure}

        The main results are summarized as follows.
	The overall evolution of the clumps in runs FE(M) and FE(L) 
	is similar to that of FE(H), and the clump lifetime is shorter
        for lower metallicity clumps.
	The mass loss rate depends on luminosity (\fref{fig:LdepFE}).
	The timescale of the 
	shock-compression phase $\tcr$ 
	is proportional to the initial ionization radius $R_{i, \rm ini}$,
	which is set by \ce{H2} photodissociation 
	and subsequent \HI~photoionization.
	The \ce{H2} photodissociation timescale 
        is estimated
	to be $\tdiss \sim 1 (\FFUV/10^3) ^ {-1}  \yr$ at the clump surface, 
	but it becomes longer in the deeper interior of the clump 
	owing to the self-shielding effect.  
	The self-shielding makes $\tdiss$ larger to
	$\tdiss \sim 3\e{4} (\FFUV/10^3) ^ {-1} (d/10^{-2}\pc)$,
	where $d$ is the depth measured from the clump surface;
	\ce{H2} photodissociation is effective only 
	for small \ce{H2} column when the FUV flux is low.
	Since EUV photons ionize the hydrogen atoms produced by photo-dissociation
	in a sufficiently short time,
	$R_{i, \rm ini}$ becomes larger with decreasing the luminosity.
	Hence, the shock-compression phase lasts longer
	in the smaller luminosity cases. 
	After the initial phase, the clumps lose mass 
	through EUV photoevaporation 
	with the mass loss rate approximately 
	given by \eqnref{eq:prate}.
	The launch velocity $v_{\rm b}$ is
	$\sim 10 \kms $, independent of the EUV luminosity.	
	The base density $n_{\rm b}$ is set by 
	the balance between ionization and recombination.
	Most of the ionizing photons are absorbed in a small volume 
	near the launch surface,
	and thus $n_{\rm b}$ is proportional to $\sqrt{\FEUV}$.
	The launching area $S$ 
	increases with decreasing the EUV luminosity,
	because the lower external pressure 
	keeps the clump volume to be large. 
	The luminosity dependence is approximately 
	expressed as $S \propto \FEUV ^{-1/3}$ during the cometary phase.
	Thus, mass loss rates due to EUV photoevaporation slightly decrease  
	 with EUV luminosity $\dot{M} \propto \FEUV ^{1/6}$.

\newcommand{\tcom}{t_{\rm com}}	
\newcommand{\megayr}{{\rm \, Myr}}	
\cite{1994_Lefloch} investigate the evolution of a neutral globule
irradiated by OB stars. They perform
2D hydrodynamics simulations with the assumption that 
	the neutral gas is isothermal.
	The lifetime of the compressed globule in the cometary phase is analytically derived as
	\eq{
		\begin{split}
		\tcom = &6.5 \braket{\frac{M_{\rm e}}{M_\odot}}^{1/3}
			\braket{\frac{\FEUV}{10^7 \unit{cm}{-2}\unit{s}{-1}}}^{-1/3}\\
			& \times \braket{\frac{T_{\rm n}}{100\Kelvin}}^{-2/3} \, {\rm Myr},	\label{eq:Lefloch}
		\end{split}
	}
	where $M_{\rm e}$ is the mass of the globule after the compression
	and $T_{\rm n}$ is the neutral region temperature. 
	Substituting the typical values of the clumps in our Z-2/FE(H) and Z0/FE(H) runs into \eqnref{eq:Lefloch},
	we obtain $\tcom = 5.6\e{4}, ~2.6\e{5}\yr$, respectively.
	The photoevaporation timescale (= clump lifetime)
	is then estimated to be $t_{\rm life} = \tcr + \tcom \simeq  6.0\e{4}\yr, ~2.6\e{5}\yr$ for 
	Z0/FE(H) and Z-2/FE(H), respectively. 
	This is in a good agreement with our simulation results 
	(\fref{fig:LdepFE}). 
	We can apply \eqnref{eq:Lefloch} to the intermediate and weak flux cases 
	to estimate the clump lifetimes,
	instead of continueing the simulations until 
	the clumps completely evaporate. 
	\begin{table}[htp]
	\caption{Adopted parameters in \eqnref{eq:Lefloch} and the estimated lifetimes.}
	\begin{center}
	\begin{tabular}{c | c c c | c} 
	Label & $\tcr \,(\yr)$ & $M_{\rm e} \,(M_\odot)$ & $T_{\rm n} \, (\Kelvin)$ & $t_{\rm life}\, (\megayr)$ \\ \hline\hline
	Z0/FE(M)&	$1.5\e{4}$	&	$0.7$	&	$10$		&	0.98	\\
	Z0/FE(L)	&	$6.0\e{4}$	&	$0.9$	&	$10$		&	4.6	\\
	Z-2/FE(M)&	$1.5\e{4}$	&	$0.7$	&	$100$	&	0.22	\\	
	Z-2/FE(L)	&	$6.0\e{4}$	&	$0.9$	&	$100$	&	1.0	\\
	\end{tabular}	
	\end{center}
	\label{tab:paramsLefloch}
	\end{table}%
	We measure $\tcr,~M_{\rm e}, T_{\rm n}$ from the simulation results 
	to estimate the lifetimes of the clumps in Z0/FE(M), Z-2/FE(M), Z0/FE(L), and Z-2/FE(L).
	The adopted values and the estimated lifetimes are tabulated 
	in \tref{tab:paramsLefloch}. 
	The resulting lifetime 
	is $\sim 5 \megayr$ with the weak luminosity for a $\metal = 1\, \smetal $ clump. 
	The corresponding EUV flux is comparable to that 
	of an early O-type star at $10\pc$ away from the clump. 
	Since such a hot star has a lifetime of a few million years \citep{2005_Stahler}, 
	distant clumps can survive, 
	although they may lose a significant fraction of the initial mass.

\begin{figure}[htbp]
	\begin{center}
	\includegraphics[clip, width = \linewidth]{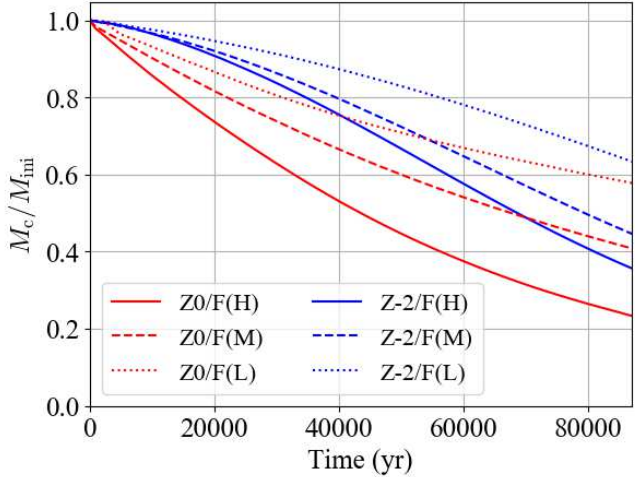}
	\caption{The time evolution of the clump mass 
	relative to the initial mass $\Mclump/M_{\rm ini}$
	for Z0/F and Z-2/F with each of the luminosity sets. 
	The solid, dashed, and dotted line indicate
	the simulations with the fiducial (H), 
	intermediate (M), and weaker (L) luminosity sets, respectively.	
	}
	\label{fig:LdepF}
	\end{center}
	\end{figure}

        The qualitative features of our runs F 
	with the intermediate and low fluxes
	are again quite similar to that with our fiducial run. 
	The solar-metallicity clumps are marginally optically thick to FUV,
	and thus photoevaporative flows are excited from the surface of the upstream hemisphere.
	The flows have velocities comparable to $c_s$,
	which is set by FUV heating. 
	Although the lower fluxes yield 
	lower temperatures of the FUV-heated region,
	the luminosity dependence is not strong;
	the temperature of the FUV-heated regions 
	varies only by a factor of three, from $100\Kelvin $ to $300\Kelvin$
	for the considered FUV luminosities. 
	Hence, the
	FUV photoevaporation rate 
	monotonically decreases with FUV flux,
	and the flux dependence is relatively small (\fref{fig:LdepF}),
	compared to the FE cases.
	For the lower metallicity $\metal < \smetal$, 
	FUV can heat the entire interior of the ``optically thin'' clumps. 
	Again, FUV heating with a higher luminosity 
	results in a slightly higher internal temperature.
	Photoevaporative flows are then driven 
	at a higher mass-loss rate
	in the higher flux cases (\fref{fig:LdepF}).

\newcommand{\tph}{t_{\rm ph}}

\subsection{The Rocket Effect and Star Formation}
	If a clump moving away from the radiation source
	survives for a sufficiently long time,
	it eventually gets out of the \HII~region
	and enters the photodissociation region (PDR).
	After the clump enters the PDR,
	its lifetime can be extended even longer.  
	In our fiducial model,
	the Str\"omgren radius of the central OB star (radiation source) 
	is estimated to be $R_{\rm S} \simeq 14 \pc \,(n_{\rm II}/10 \cm{-3})^{-2/3}$, 
	where $n_{\rm II}$ is the average density of the \HII~region,
	with neglecting dust absorption of EUV.
\footnote{
	The Str\"omgren radius can be reduced by $\sim 20\, \%$ owing to
	dust absorption \citep{1978_Spitzer}.
}
	The timescale where the accelerated clump gets out of the \HII~region is calculated as
\newcommand{\tout}{t_{\rm out}}
	\eqarray{
	\tout &=& \frac{ R_{\rm S}} {b} \frac{2}{  1 + \sqrt{1 + 4aR_{\rm S}/b^2}  } \label{eq:tout} \\
	&\sim& 0.63 \, \left[{\left(\frac{n_{\rm II}}{10 \cm{-3}}\right)^{-1/3} } - 0.17\right] \megayr .	\label {eq:appro5}
	}
	Here we have assumed $n_{\rm HII} \lesssim 10^3 \cm{-3}$.
	In order to compare $\tout$ 
	with the lifetime of a photoevaporating clump $\tph$, 
	we apply the clump model of \eqnref{eq:rocket}
	to estimate $\tph$,
	\eq{
	\begin{split}
		\tph  = &
		\frac{M_{\rm cl}}{ S \rho_{\rm b} v_{\rm b}}\\
		\sim & 0.40
	\left(\frac{S}{10^{-4} \pc ^2}\right)^{-1}
	\left(\frac{n_{\rm b}}{10^5 \cm{-3}}\right)^{-1}	\\
		& \times \left(\frac{v_{\rm b}}{10 \unit{km/s}{}}\right)^{-1}
	\left(\frac{M_{\rm cl}}{M_\odot } \right) 
				 \megayr,
	\end{split}
	}
	for the solar-metallicity case.
	Hence, the photoevaporation time $\tph$
	is comparable to $\tout$ in our model;
	clumps could either evaporate 
	before reaching the PDR, or 
	enter the PDR with small masses.
	Overall, 
	solar-metallicity clumps  
	lose most of the mass while they are in the \HII~region,
	and thus star formation 
	can be suppressed around massive stars. 

	The clumps which survive to go out of the \HII~region 
	are still exposed to the FUV radiation field in the PDR. 
	Since the compression in the \HII~region 
	makes an optically-thick dense clump to FUV ($\eta_0 \sim 2\e{2}$), 
	the evolution of the compressed clump 
	would be different from that of the F case presented in Section 4.
	\cite{2002_Gorti} show that 
	the evolution of clumps with $\eta_0 > 1$ bifurcates,
	depending on the ratio of the sound speed in the PDR to 
	that in the cold clump $\nu \equiv c_{\rm PDR}/c_{\rm cl}$.
	If $\eta_0 < 4 \nu^2 / 3 (\equiv \eta_{\rm crit})$, 
	a PDR ``shell'' is produced in the cold clump, where the high pressure drives 
	a shock and compresses the clump. 
	The compression continues until the column density of the clump increases to $\eta_{\rm crit}$.
	If $\eta_0 > \eta_{\rm crit}$, 
	a thinner PDR shell is produced compared to the case with $\eta_0 < \eta_{\rm crit}$,
	and the shell expands at $\sim c_{\rm PDR} $. 
	The thermal pressure inside the shell rapidly declines below
	the cold clump pressure because of the expansion.
	The clump expands
	until the column density decreases to $\eta_{\rm crit}$,
        and then it keeps
	the constant column density of $\eta_{\rm crit}$.
         The clump shrinks in size because of the mass loss due to photoevaporation,
        and thus it gets denser to yield the constant column density. 
	If the density could get sufficiently high, 
	star formation would take place in the surviving clump \cite{2002_Gorti}.

\subsection{Effects of Gravity}
	\label{sec:gravity}
	Self-gravity is ineffective to prevent clumps from photoevaporating. 
        In fact, at the launching points of EUV-driven flows, 
	the free fall time
	\eq{
		\tff  = \sqrt{\frac{3\pi}{32 G \rho}} \sim 0.2\,
		\left(\frac{\nh}{10^5\cm{-3}}\right)^{-1/2} \megayr \, ,
	}	
	is much longer than the crossing time of
	the photoevaporative flows $\tcr$.
	This indicates that 
	EUV-driven photoevaporation occurs 
	even if self-gravity were incorporated in our simulations.
	Several previous studies conclude, by performing
        three-dimensional simulations with gravity,
	that the gas self-gravity does not change the overall evolution
	\citep{2003_Kessel-Deynet, 2007_Esquivel}. 
	
	Nevertheless, 
	since the surface area is determined by the internal structure of a clump,
	incorporating gravity can affect the photoevaporation rate and the clump lifetime.
	The typical ratio of the gravitational energy to the thermal energy 
	for the clump gas is
	\eqarray{
		\psi &=& \frac{GM_{\rm cl}}{R_{\rm cl} c_s^2} \\
			&\sim & 5 \left(\frac{\Mcl}{M_\odot}\right)
			\braket{\frac{\Rcl}{10^{-2}\pc}}^{-1}
			\braket{\frac{T}{10\Kelvin}}^{-1}	\, .	
	}
	The ratio $\psi$ is of the order of $10^{-3} - 10^{-2}$
	for low-metallicity clumps with $\metal \lesssim 10^{-2} \, \smetal$
	whose typical size and temperature are $\Rcl \sim 0.1\pc$ 
	and $T \sim 10^3 - 10^4 \Kelvin$, respectively.
	Therefore, 
	the lifetimes of the low-metallicity clumps would not be significantly affected 
	even if we take account of gravity (\fref{fig:lifetimes}).
	On the other hand, $\psi$ is of the order of unity 
	for the clumps with $\metal \gtrsim 10^{-1} \, \smetal$;
	self-gravity may reduce the clump radius.
	Although the density of the clump center can increase,
	the base density is physically set by the EUV flux
        and the recombination reaction coefficient,
	and thus independent of whether or not the clump contracts gravitationally.
	Hence, the smaller surface area of the higher metallicity clumps
        reduced by the effects of gravity
	allows them to survive even longer.

        Regarding FUV photoevaporation, 
	the crossing time for FUV-driven flows are sufficiently shorter 
	than the free fall time for the clumps with $\metal \gtrsim 10^{-1.5} \, \smetal$;
	FUV photoevaporation can be excited even if the gravity is incorporated. 
	The ratio $\psi$ is small for the higher metallicity clumps,
	and thus the lifetimes of the higher metallicity clumps 
	would not be significantly changed 
	owing to the effects of gravity.
	In contrast, 
	the crossing time is comparable to the free fall time for 
	the lower-metallicity clumps with $\metal \lesssim 10^{-2} \,\smetal$.
	Incorporating gravity 
	would suppress the dispersal of the clumps. 
	Especially for $\metal = 10^{-3} \, \smetal$, 
	the clump is likely to lose its mass owing to the neglect of gravity in our model.

	In addition to the suppression of photoevaporation,
	self-gravity is essential for collapse of clumps to form stars inside
	\citep[radiation-driven collapse;][]{2007_Esquivel}.
	The free fall time of the highest density regions in the cometary clumps 
	is of the order of $10^4 \yr$,
	which is shorter than the lifetimes of clumps with $\metal \gtrsim 10^{-1.5} \, \smetal$. 
	Star formation may occur in such clumps while the parental clumps evaporate.
	This directly sets the star formation efficiency around massive stars;
	it is worth investigating collapse of clumps illuminated by external massive stars
	in future hydrodynamics simulations.

\section{Summary \& Conclusion}	\label{sec:summary}
We have performed a suite of 3D radiation hydrodynamics simulations
with nonequilibrium chemistry 
to study photoevaporation of clumps
exposed to UV radiation from massive stars.
We have derived the lifetimes of the clumps 
with a wide variety of metallicities
$10^{-3} \, \smetal \leq \metal \leq \smetal$.

In our fiducial model,
the clumps exposed to the EUV radiation 
are shock-compressed for the first $\tcr \simeq 4 \times 10^3 \yr$.
About a half of the initial mass is lost by photoevaporation
in this early phase.
After the shock-compression and implosion phase ends, 
the mass loss rate is determined essentially by the surface area
of the clump. Since lower-metallicity clumps have higher internal
temperatures, their volume remain large and lose mass faster.
Consequently, the lifetime is shorter for lower-metallicity clumps 
in the range of $10^{-2} \, \smetal \leq \metal \leq \smetal$,
and scales approximately as $T_{\rm life}  = [9.6 + 3.6 \log (\metal / \smetal)] \e{4} \yr$.
The clumps with $\metal \lesssim 10^{-2} \, \smetal$
keep nearly the same volume after the compression phase,
and thus the lifetime is almost independent of metallicity
with $T_{\rm life} \simeq 10^4\yr$. 

In order to study clump photoevaporation 
in the photodissociation regions, 
we have also performed a set of simulations where 
the EUV radiation is disabled.
The simulations (dubbed F runs) show that the clumps
with $\metal \simeq \smetal $
are marginally optically thick to FUV photons,
which heat only the hemisphere facing 
the external radiation source. 
The clumps with $\metal < \smetal$ are optically thin to FUV photons
because of the reduced amount of dust.
Since the FUV heating rate increases with metallicity,
clumps with higher metallicity evaporate faster.
However, when the metallicity is close to $1 \, \smetal $,
the photoevaporation rate is actually {\it smaller} 
than the sub-solar clumps because photoevaporative winds are
launched only in the hemisphere.
The clump lifetimes in the F runs are generally longer
($\gtrsim 10^5\yr$) than those in the FE runs.

Clumps are accelerated to move away from the radiation source 
by the rocket effect. 
For a clump with $\metal = 1\, \smetal$,
the typical velocity is $\sim 10 \unit{km}{}\unit{s}{-1}$ 
when EUV field is present, 
while the velocity is much smaller to be $\sim 1\unit{km}{}\unit{s}{-1}$ without EUV
radiation. 
For low-metallicity ($\metal < \smetal$) clumps exposed to EUV,
the receding velocity is $\sim 10 \unit{km}{}\unit{s}{-1}$. 
The larger launch area of the lower metallicity clump 
yields a more efficient acceleration.  

In the local, present-day star-forming regions with $\metal \gtrsim 10^{-1}\, \smetal$,
star formation can be promoted in the \HII~region around massive stars
because of the density enhancement due to the radiation-driven implosion.
On the other hand, in the PDRs, 
FUV-heated clumps expand 
and thus star formation can be likely delayed or suppressed. 
In low-metallicity environments with $\metal \lesssim 10^{-2}\, \smetal$, 
the clumps are not compressed in the \HII~region, 
and have short lifetimes.
Hence star formation can be significantly suppressed.
In the PDRs,
inefficient FUV heating allows clumps 
to survive for a longer time than the free fall time.

\section*{Acknowledgement}
We thank Takashi Hosokawa, Kazumi Kashiyama,
Anastasia Fialkov, and Tilman Hartwig 
for fruitful discussions and insightful comments on the paper. 
RN has been supported by the Grant-in-aid for the Japan Society for the
Promotion of Science (16J03534) and by Advanced Leading Graduate Course 
for Photon Science (ALPS) of the University of Tokyo. 
All the numerical computations were carried out on Cray XC50 at Center
for Computational Astrophysics, National Astronomical Observatory of Japan.

\bibliographystyle{../apj}
\bibliography{../template}
\end{document}